\documentclass[10pt,letter]{IEEEtran}
\usepackage{graphicx}
\usepackage{subfigure}

\usepackage[T1]{fontenc}
\begin{document}


\title{
  An Isolated Power Factor Corrected Power Supply Utilizing the
  Transformer Leakage Inductance.
}
\author{ Thomas Conway 
\thanks{
Thomas Conway is a lecturer in the ECE dept. of the University of Limerick, Limerick, Ireland
}
\thanks{Contact: 
Dr. Thomas Conway
Lecturer, ECE Dept
University of Limerick
National Technology Park
Limerick, Ireland.
Tel +353 61 202628,
Email thomas.conway@ul.ie
}
\thanks{
This work has been submitted to the IEEE for possible publication.
Copyright may be transferred without notice, after which this 
version may no longer be accessible.
}

}

\maketitle

\begin{abstract}
The widespread use of electronic devices increases the need for compact
power factor corrected power supplies. This paper describes an isolated
power factor corrected power supply that utilizes the leakage inductance
of the isolation transformer to provide boost inductor functionality.
The bulk capacitor is in the isolated part of the power supply allowing
for controlled startup without dedicated surge limiting components. 
  A control method based on switch timing and input/output voltage 
measurements is developed to jointly achieve voltage regulation and
input power factor control.
A prototype design is implemented with detailed measurements and waveforms
shown to confirm the desired functionality.
\end{abstract}


\section{Introduction}

The widespread use of electronic devices from single phase AC supplies
necessitates the increasing use of power factor corrected (PFC) power supplies
in many applications including electronic equipment, computer servers 
and consumer products.  PFC power supplies provide low total harmonic distortion
(THD) in the current drawn from the line and this is an increasingly important
requirement. 

Power factor correction techniques have been researched
widely in the literature \cite{ref:pfcgeneral}\cite{ref:rev2} and active 
PFC using high frequency switching techniques \cite{ref:rev1} are now commonly
used.  The overarching principle involves controlling the input current drawn
from the mains input to achieve the required current shape for low THD and high
power factor.  The power supply must provide a regulated DC output voltage 
and for many applications, galvanic isolation is also required.

The basic boost or step-up converter \cite{ref:MohanBook} forms the core of most
architectures as it has an input inductor that allows input current control
to be readily achieved.  The well known flyback converter can be derived from the
buck-boost converter, but with a transformer for output voltage isolation \cite{ref:MohanBook}.

Traditionally for PFC supplies, flyback converters 
have been used for lower power levels ($\le $ 100W ).  For higher power levels
($\ge $ 500W), a separate boost converter for PFC and separate DC to DC converter
with transformer isolation for output DC voltage regulation is used.
\subsection{Brief Review of Published Isolated PFC Converters}
For lower power levels, flyback type architectures, often using a 
single switching element can provide PFC functionality,
and use an output bulk capacitor for energy storage.  
A range of flyback based power factor corrected power supplies have
been developed and are described in the literature. Ref.~\cite{ref:pfcflyback}
describes a 60W flyback PFC supply to achieve IEC61000 THD requirements.
A 10W to 30W LED lighting supply is described in 
\cite{ref:pfcflybackLED} and another 60W supply described in \cite{ref:pfccmp1}.
A 72W flyback design is shown in \cite{ref:pfccmp3} and a 100W 
flyback PFC design in described in \cite{ref:pfcfly100W36V}.
In these architectures,
the flyback transformer initially stores energy from the input source
and then releases it to the output bulk capacitor.  The transformer
thus provides the PFC input current control and galvanic isolation.
However, the need to store all the energy in the flyback transformer,
the unidirectional core excitation \cite{ref:MohanBook},  high voltage stresses
in the switching device and difficulties with the transformer leakage inductance, 
limit the usage of such power supplies to lower power levels.

At higher power levels, two stage supplies with a separate boost PFC stage and
isolated DC to DC stage are widespread.
They typically consist of a power factor correction stage, 
based around a boost converter with large bulk storage capacitor 
to control the input line current, followed by 
an inverter driving a high frequency isolation transformer and finally an output stage 
with at least some filtering components.  Such designs work well, especially
when there is a need to supply a number of different voltage rails and a 
long holdup time is required.  
Most of these architectures use one (or more) boost inductors in the active 
PFC stage \cite{ref:rev1},
and such inductors need to handle the full supply power levels resulting
in a significantly sized component.   The design of such front end PFC 
converters is a tradeoff between boost inductor size and high frequency 
losses \cite{ref:PostxfmrCbulkandBoostInductorSize}.
Also, cascading multiple power stages leads to reduced overall efficiencies.
The front end boost PFC stage typically has  high in-rush
current on startup, or needs additional components to limit the in-rush
current \cite{ref:pfcinrush}\cite{ref:pfcinrush2}.

Therefore, there has been considerable research published on other architectures
that attempt to eliminate or mitigate some of the disadvantages of both the flyback
and two stage PFC AC/DC power supplies.

A quasi-active PFC converter in \cite{ref:pfccmp5} delivers 100W using a combined PFC cell
with two transformers, one operating as a flyback converter and one operating as a  forward
converter.  This allows parallel power transfer and avoids  lossy snubber networks, all with
a single controlled switch, but at the expense of a number of magnetic cores and inductor elements.

A combination of a separate boost stage cascaded with a lossless
snubber network is presented in \cite{ref:pfccmp2}, delivering 96W.  Again only a single controlled switch
is used but a number of separate magnetic cores and inductor elements are required.

For higher power levels, modified two stage structures have been proposed.
An architecture to improve on the existing two stage  PFC is described 
in \cite{ref:PostxfmrCbulkandBoostInductorSize} which 
moves the boost inductor and bulk capacitor to the isolated side of
a resonant LLC converter with the effect of improving the 
voltage stresses on the switching devices and providing an output power of 250W.

A two stage structure proposed in \cite{ref:pfccmp6} uses a boost converter PFC with
a resonant LLC converter. The design uses transformer coupled power transfer from
the boost inductor to the output. Using two transformers (one providing the boost inductor 
functionality), a power output of 480W is provided.

\subsection{Proposed PFC Architecture}
In this paper an active power factor corrected power supply is described,
whereby the leakage inductance of the high frequency isolation transformer is used
to provide the functionality of the boost inductor.  
Minimization of the leakage inductance in high frequency isolation transformers
is normally desirable in most DC to DC converters, although resonant and soft switching
architectures do use a controlled amount of leakage inductance \cite{ref:cntlleakageinduct} 
for the purpose of reducing switching losses.

The use of a controlled amount of leakage inductance is proposed in this paper
to eliminate the need for two separate magnetic components in the two stage 
PFC converter and instead uses one magnetic component to achieve both the power 
factor correction and galvanic isolation.  

In-rush current on startup can also be controlled by implementing a soft
start strategy whereby the large bulk capacitor is initially charged up in
a controlled manner.

Bidirectional core excitation is used, with part of the energy
transferred via transformer action, and part stored in the 
transformer leakage inductance.  The described architecture provides
a useful technique at power levels above those suitable for single stage
flyback type converters.

The technique lends itself to the adoption of wide band-gap semiconductor devices \cite{ref:GANgeneral}
with hard switching \cite{ref:SiCHardSwitched}\cite{ref:SiCHardSwitched2}.
Typically applications might include LED lighting, electronic equipment, 
server power supplies and on-board chargers for electric vehicles.

In this paper,  section~\ref{sec:proposedarch} describes the proposed 
architecture.
A simplified system model for the power electronics is outlined in 
section~\ref{sec:simpsysmodel}, from which the basic operating modes 
of the circuit are identified and
characterized.  
The dynamic selection of the operating modes and their control
parameters to achieve power factor correction are presented 
in section~\ref{sec:psucontrol}.  
An example design for a 300W 50V power supply is shown
in section~\ref{sec:egdesign} with an experimental prototype 
constructed, its operation confirmed, and waveforms and measurements presented
 in section~\ref{sec:prototypemeasure}.


\begin{figure*}[htb]
  \centering
    \includegraphics[width = 0.75\textwidth]{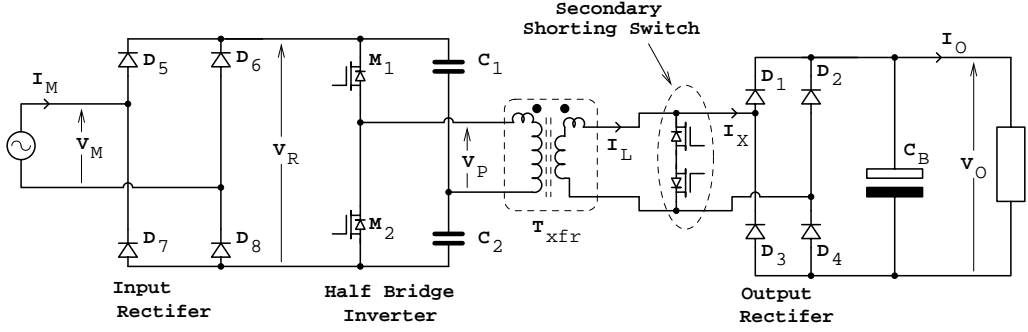}
  \caption{Circuit diagram of the proposed power supply architecture.}
  \label{fig:principle}
\end{figure*}
\section{Proposed Architecture\label{sec:proposedarch}}
The circuit diagram of the proposed power supply is shown in 
Fig.~\ref{fig:principle}. 
A conventional 4 diode full wave rectifier
rectifies the input AC source voltage producing a voltage $V_R$.
This voltage is inverted to the high frequency $f_s$ with a half
bridge inverter before being applied to a high frequency transformer
$T_{xfr}$. 
  The half bridge inverter consists of the two switches M1 and M2
  operated out of phase with a 50\% duty cycle at the switching
  frequency $f_s$ and the capacitive divider formed by C1 and C2.
  The capacitors C1 and C2 prevent DC current flowing through the
  transformer primary and causing saturation problems.
  The values of C1 and C2 are chosen sufficiently small such that 
  at the mains frequency $f_{AC}$ and low power level, they
  allow the rectifier output voltage $V_R$ to follow the input
  mains waveform envelope. However, at the switching frequency, their
  values are sufficiently large to act as fixed voltage sources
  and not resonate with the transformer inductances or load.

\begin{figure}[htb]
  \centering
        \includegraphics[width=0.5\textwidth]{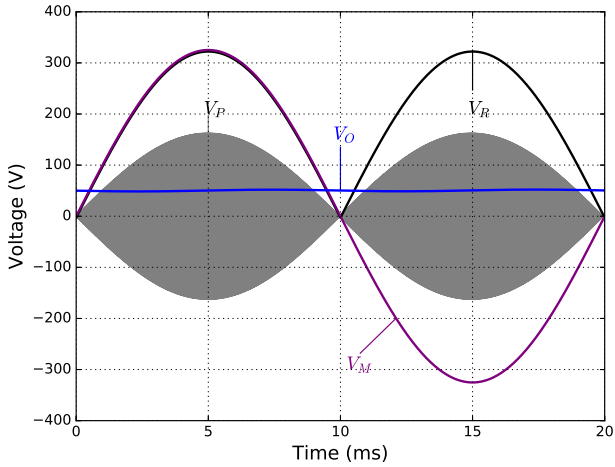}
  \caption{Qualitative voltage waveforms for the circuit of Fig.~\ref{fig:principle} over a full AC cycle.}
  \label{fig:idealvoltages}
\end{figure}

\begin{figure}[htb]
  \centering
        \includegraphics[width=0.5\textwidth]{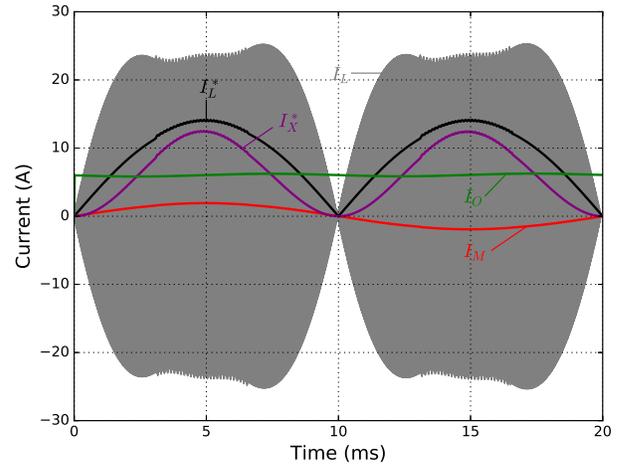}
  \caption{Qualitative current waveforms for the circuit of Fig.~\ref{fig:principle} over a full AC cycle.}
  \label{fig:idealcurrents}
\end{figure}

Fig.~\ref{fig:idealvoltages} and Fig.~\ref{fig:idealcurrents} show 
qualitative voltage and current waveforms for the circuit of 
Fig.~\ref{fig:principle}.
The mains input voltage is $V_M(t) = \sqrt{2} V_{AC} \sin (2\pi f_{AC} t)$, with
$V_{AC}$ being the rms input voltage and $V_R(t)$ being the input 
voltage fully rectified.  The transformer
primary voltage $V_P(t)$ switches at the high frequency rate $f_s$, 
but with an amplitude of half $V_R(t)$, due to the half bridge configuration.

The symbol for the transformer in Fig.~\ref{fig:principle}
is drawn to emphasize that the transformer leakage inductance is 
used in the circuit rather than the usual case whereby leakage inductance
is minimized as much as possible.  The key to the operation of
the circuit is the bidirectional secondary shorting switch shown
in Fig.~\ref{fig:principle} \cite{ref:SoftSwitchDCDC}.

By turning on this switch at the beginning of the high frequency cycle,
the current in the leakage inductance of the transformer is controlled
by the input voltage level and the period that the switch is on for (denoted
$T_1$).

When the shorting switch opens, the leakage inductance current is
forced through the output rectifier into the bulk capacitor $C_B$ 
and output load.
The bulk capacitor $C_B$ stores sufficient energy over the input 
AC half cycle to provide a low ripple output voltage $V_O(t)$.

The output voltage $V_O(t)$, should not be less than the maximum value of
$V_P {N_S\over N_P}$ where $N_S\over N_P$ is the transformer turns ratio.
In Fig.~\ref{fig:idealcurrents}, the current waveforms labeled 
$I_L^*$ and $I_X^*$ represent the currents $I_L(t)$ and $I_X(t)$
(as in Fig.~\ref{fig:principle}), 
multiplied by the sign of the primary voltage $V_P(t)$
and averaged over each high frequency period $T$.  The value of $I_L^*$
is the same shape as the mains input voltage and thus the
magnitude of the filtered line input current $I_M(t)$ equals 
$ {1\over 2}I_L^*(t) {N_S\over N_P} $.

Provided the DC output voltage $V_O$ is larger than the maximum secondary
output voltage, 
the leakage inductance current is controlled by the input 
voltage level, the output DC voltage level and the secondary shorting switch
timing period $T_1$. 
With measurement of the voltage levels, the timing period $T_1$
is used to control the input current $I_M(t)$ drawn from the AC source and
thus power factor correction can be achieved.

In practice,
this current control functionality operates at a high switching frequency $f_s$ and 
controls the average input current over the switching  period $T = {1\over f_s}$.
Output voltage regulation is achieved by controlling the ratio of 
instantaneous current drawn to the instantaneous voltage applied and this control
functionality operates at a slow rate comparable with the input source frequency, denoted $f_{AC}$.

In effect, the transformer leakage inductance and secondary shorting switch 
act as a step up or boost converter with the output rectifier and 
bulk capacitor $C_B$.  However, the input voltage polarity reverses once
each cycle and the boost operation is carried out in two $T/2$ time periods, 
each with opposite polarity voltage and current waveforms.

\section{Theory of Operation\label{sec:simpsysmodel}}
\begin{figure}[htb]
  \centering
    \includegraphics[width = 0.5\textwidth]{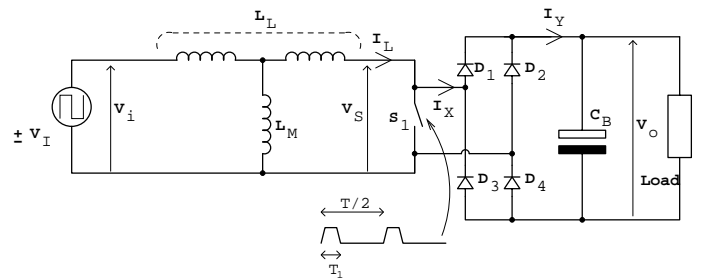}
  \caption{Simplified circuit model for the proposed power supply.}
  \label{fig:model}
\end{figure}

Fig.~\ref{fig:model} shows a simplified circuit model for the proposed power supply.
Assuming the switching frequency of the converter is very high compared to to the
AC source frequency, the input to the transformer can be considered to be 
essentially a 50\% duty cycle square wave with period $T$ and peak amplitude $\pm V_I$. 
The model in 
Fig.~\ref{fig:model} is referenced to the secondary side of the transformer 
and the voltage amplitude into the transformer model is the primary 
voltage $V_P(t)$  multiplied by the turns ratio of the transformer, or
\begin{equation}
  \label{eqn:VIscaling}
	V_I[kT] =\left| V_P(kT) {N_s \over N_p} \right |
\end{equation}
at time $t = k T$.

The operation of the circuit is based on the assumption that the transformer magnetizing 
inductance $L_M$ has little effect on the operation of the circuit other than to add a
magnetizing current to the input source.  Simulations show that magnetizing currents 
significantly less (a factor of ${1\over 5}\times$ or less), than the currents being 
transferred, 
have little impact on the circuit overall functionality.  This is verified in the experimental
prototype waveforms of section~\ref{sec:prototypemeasure}.

The total leakage inductance of the transformer is
denoted $L_L$ and the current flowing out of the transformer secondary winding
is denoted $I_L(t)$.

The operation of the system is essentially that of a step up or 
boost converter and is based around the timing of shorting switch $S_1$ in 
Fig.~\ref{fig:model}.  At the beginning of a switching cycle, the input voltage
switches to $+V_I$ (dropping the $[kT]$ for notation for clarity) and simultaneously
the shorting switch $S_1$ is turned on. 
The current $I_L(t)$ in the leakage 
inductance $L_L$ rises linearly while the switch $S_1$ is on.  
  When the switch $S_1$ is turned off, the current in the leakage inductance
is forced through the rectifier diode bridge formed by $D_1$, $D_2$, $D_3$, and $D_4$,
and into the capacitor $C_B$ and system load, and the current in the leakage inductance falls.  
After a period of $T\over 2$, the input voltage changes sign to $-V_I$
and the same operation occurs, except for a change in the sign of the inductor 
current.
  Two distinct operation modes of the circuit can be identified depending on
whether the leakage inductance current starts at zero and returns to zero
before time $T\over 2$, denoted as the discontinuous conduction mode (DCM), or when the
leakage inductance current starts the cycle with a non zero (negative) value,
retains a non zero (positive value) at time $T\over 2$ and returns to a 
non zero (negative) value at the end of the cycle (time $T$), denoted as the 
continuous conduction mode (CCM).  

  To achieve unity power factor, the circuit needs to be operated in such a manner
as to control the input current drawn from the supply.
The two operating modes are now discussed in detail to relate the input current 
drawn to the timing period $T_1$.

\subsection{Discontinuous Conduction Mode (DCM) \label{secDCM}}

\begin{figure}[htb]
  \centering
    \includegraphics[width = 0.4\textwidth]{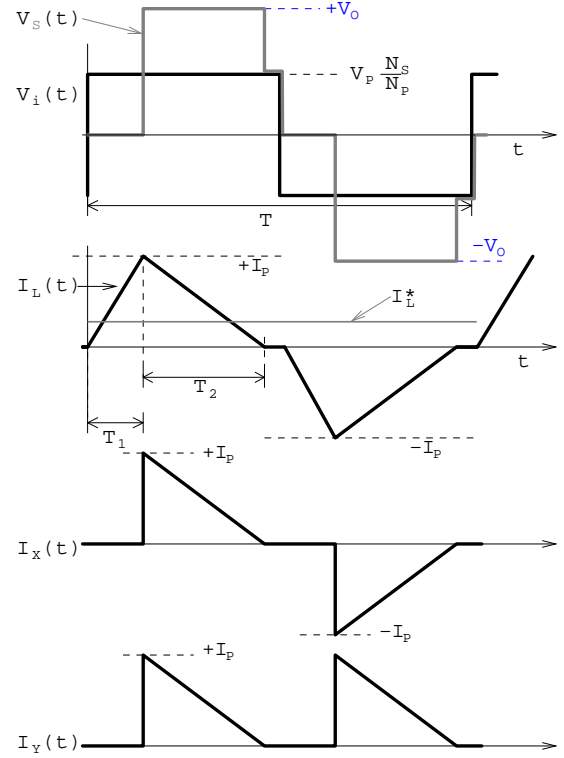}
  \caption{Idealized discontinuous conduction mode waveforms.}
  \label{fig:DCMmode}
\end{figure}

Fig.~\ref{fig:DCMmode} shows the input voltage $V_i(t)$, the secondary voltage $V_S(t)$, the  leakage inductor
current $I_L(t)$ and the current into and out of the output rectifier $I_X(t)$ and $I_Y(t)$, 
for the circuit operating in discontinuous conduction mode.  
With the shorting
switch $S_1$ closed, the 
leakage inductor current $I_L(t)$ rises from zero to the value $+I_P$ over
the set period $T_1$, thus:
\begin{equation}
 \label{eqnDCM1}
  I_P =  V_I T_1 / L_L
\end{equation}
When the shorting switch $S_1$ opens, the inductor current falls back to zero 
over a period $T_2$ with the relationship:
\begin{equation}
 \label{eqnDCM2}
  I_P =  (V_O-V_I) T_2 / L_L
\end{equation}

The sum of the periods must be less than the half period $T/2$ to ensure
operation in the discontinuous conduction mode or:
\begin{equation}
 \label{eqnDCMCCMboundary}
  T_1 + T_2 \le {T\over 2} .
\end{equation}
The average input current to the transformer model (ignoring the magnetizing
inductance) over the period $T/2$ can then be calculated as:
\begin{equation}
  I_L^* = {1\over 2} I_P { T_1 +T_2 \over {T\over 2} }
\end{equation}
and combining with eqn.~\ref{eqnDCM1} and eqn.~\ref{eqnDCM2}, the average
input current is:
\begin{equation}
  \label{eqn:IAcalcDCM1}
  I_L^* = {T_1^2 \over T L_L}\left({V_I V_O \over V_O- V_I}\right)
\end{equation}
The actual input current from the AC source is a scaled version of this
current and is: 
\begin{equation}
  \label{eqn:IADCM1}
  I_M =  {1\over 2} {N_s \over N_p} I_L^*
\end{equation}
with any contribution from the magnetizing inductance averaging to zero
over each $T$ period.

It is apparent by considering eqn.~\ref{eqn:VIscaling} and 
eqn.~\ref{eqn:IADCM1}, that achieving unity power factor in the
input source is equivalent to controlling the current value $I_L^*$ to be
directly proportional to $V_I$.  Denoting the constant of proportionality
as $G_M$, or $ I_L^* = G_M V_I$, then substituting in eqn.~\ref{eqn:IAcalcDCM1}
and rearranging yields the equation:
\begin{equation}
  \label{eqn:T1calcDCM1}
  T_1 = \sqrt{
    G_M T L_L \left({V_O -V_I\over V_O}\right)
  }.
\end{equation}
The equation shows that given a constant of proportionality
as $G_M$, the required time period $T_1$ can be calculated
by knowledge of the system parameters $L_L$ and $T$,
measurement of the output voltage $V_O$ and calculating
$V_I$ by measurement of the rectified input source voltage
and scaling by a factor of ${ 1 \over 2 } {N_s \over N_p}$.

\subsection{Continuous Conduction Mode (CCM)\label{secCCM}}

\begin{figure}[htb]
  \centering
    \includegraphics[width = 0.4\textwidth]{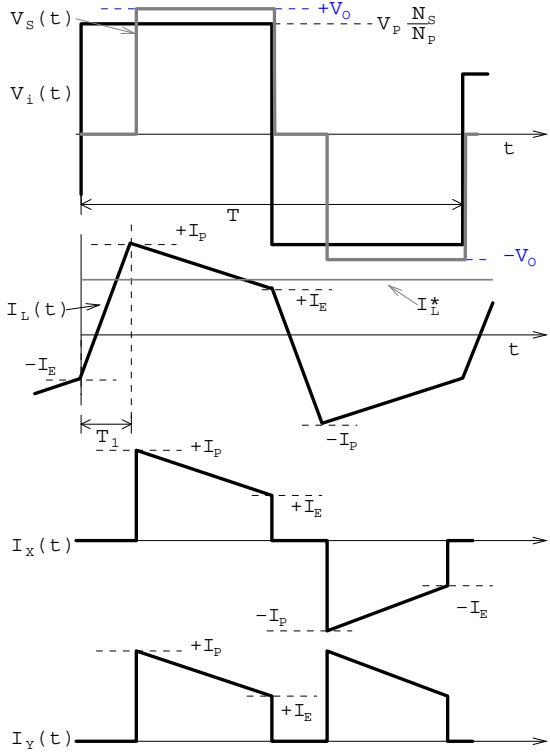}
  \caption{Idealized continuous conduction mode waveforms.}
  \label{fig:CCMmode}
\end{figure}

Fig.~\ref{fig:CCMmode} shows the input voltage $V_i(t)$, the secondary voltage $V_S(t)$, the 
leakage inductor current $I_L(t)$ and the current into and out of the output rectifier $I_X(t)$ and $I_Y(t)$,
for the circuit operating in continuous conduction mode.
 With the shorting switch $S_1$ closed, the
leakage inductor current $I_L(t)$ rises from the value $-I_E$  to the 
value $+I_P$ over the set period $T_1$ thus:
\begin{equation}
 \label{eqnCCM1}
  I_P +I_E  =  V_I T_1 / L_L
\end{equation}
When the shorting switch $S_1$ opens, the inductor current falls back to $+I_E$
over a period $T_2 = {T\over 2} - T_1$ with the relationship:
\begin{equation}
 \label{eqnCCM2}
  I_P-I_E =  (V_O-V_I) T_2 / L_L
\end{equation}
The average input current to the transformer model (ignoring the magnetizing
inductance) over the period $T/2$ can then be calculated as:
\begin{equation}
  I_L^* = {  {1\over 2} (I_P -I_E) T_1 + {1\over 2} (I_P +I_E) T_2 
     \over {T\over 2} }
\end{equation}
and combining with eqn.~\ref{eqnCCM1} and eqn.~\ref{eqnCCM2}, the average
input current can be shown to be:
\begin{equation}
  \label{eqn:IAcalcCCM1}
  I_L^* = \left( T_1 {T\over 2} - T_1^2 \right) { V_O \over T L_L}
\end{equation}
With $ I_L^* = G_M V_I$, then substituting in eqn.~\ref{eqn:IAcalcCCM1}
and rearranging yields the equation:
\begin{equation}
  \label{eqn:T1calcCCM1}
  T_1 = {T\over 4} \left(
   1 - \sqrt{
           1 - { 16 G_M L_L V_I \over T V_O}
            } \right)
\end{equation}
Similar to section~\ref{secDCM}, this equation shows that 
given a constant of proportionality $G_M$, the required time 
period $T_1$ can be calculated from $L_L$, $T$, $V_O$ and $V_I$.

\subsection{Power Handling Capability\label{sec:powercap}}

The power capability of converter is determined by the maximum value of $G_M$ supported
which is limited by the requirement for equation~\ref{eqn:T1calcCCM1} to result 
in a real number.  This requires the argument under the square root to be non-negative
and hence 
$$
  { 16 G_{Mmax} L_L V_I \over T V_O} \le 1
$$
and with a maximum input voltage $V_{Imax} = \sqrt{2} V_{AC} { 1 \over 2 } {N_s \over N_p}$, 
yields a power capability $P_{max} = {1\over 2}G_{Mmax} V_{Imax}^2$ of:
\begin{equation}
  \label{eqn:maxpowercalc}
    P_{max} \le { V_{AC}  {N_s \over N_p} V_O \over 32 \sqrt{2} f_s L_L }
\end{equation}
This equation can be used as a basis for converter design as demonstrated
by the prototype example in section~\ref{sec:egdesign}.
The maximum peak current in the leakage inductor during the CCM can be calculated
as:
\begin{equation}
  \label{eqn:Ipeakcalc}
    I_{Pmax} = { V_O T \over 8 L_L }
\end{equation}
and the transformer must be designed to handle this peak current without
saturation.


\section{Power Supply Control\label{sec:psucontrol}}

\begin{figure}[htb]
  \centering
    \includegraphics[width = 0.5\textwidth]{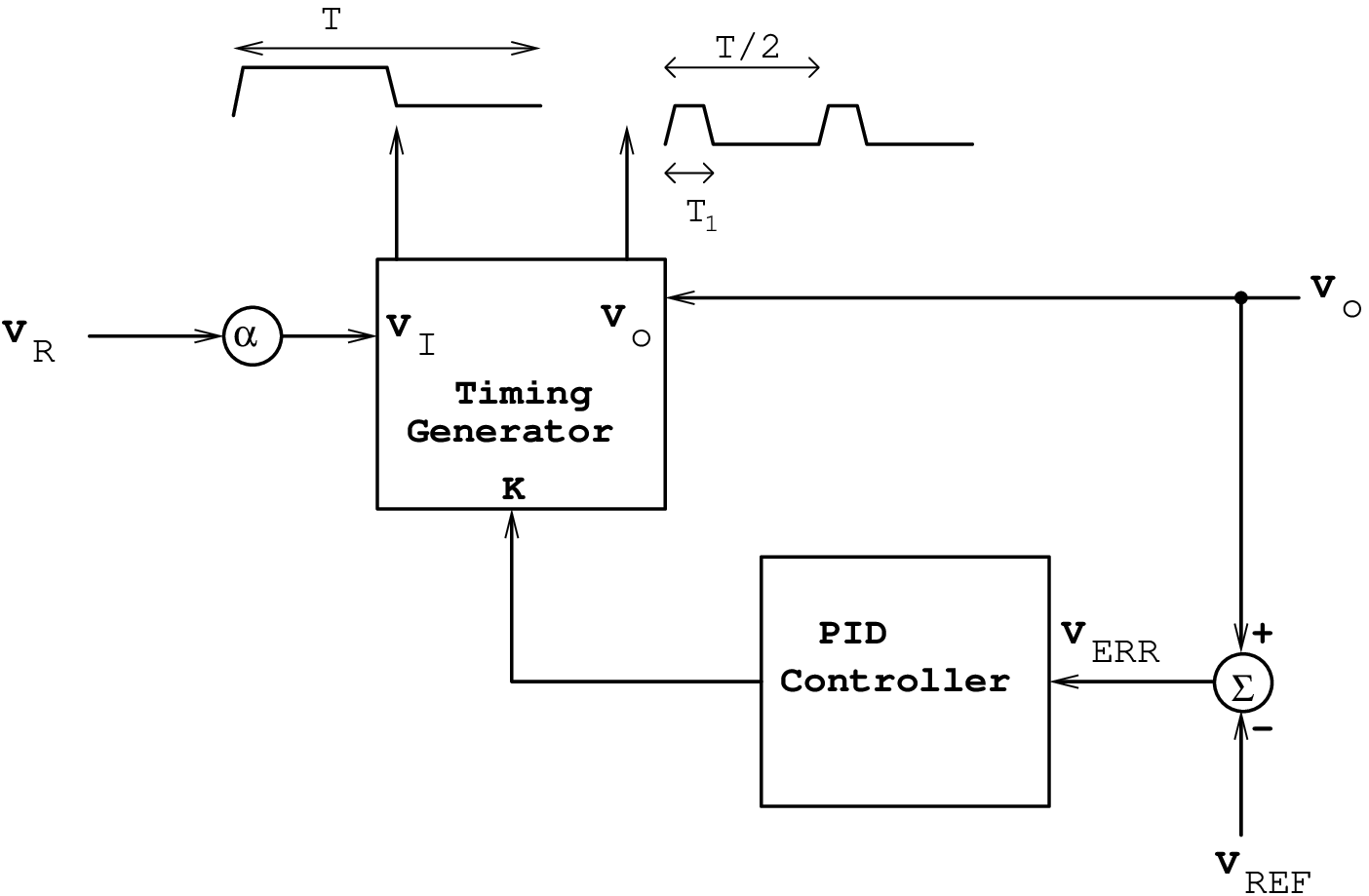}
  \caption{Feedback control for the power supply.}
  \label{fig:feedback}
\end{figure}

The control objective for the power supply is to provide a constant output
voltage and unity input power factor.  This requires measurement of the
output voltage and adjustment of the input current through the $G_M$ factor
defined in section~\ref{secDCM}.  However, calculating the time parameter
$T_1$ in section~\ref{secDCM} and ~\ref{secCCM} also requires knowledge of
the parameter $L_L$, the leakage inductance, which may not be accurately
known.  Therefore a new control parameter $K$ is defined as:
\begin{equation}
  K = { G_M L_L \over T}
\end{equation}
and $K$ is used for control rather than $G_M$.
Substituting into eqn.~\ref{eqn:T1calcDCM1} and eqn.~\ref{eqn:T1calcCCM1}
results in the required calculations for DCM as:
\begin{equation}
  \label{eqn:T1calcwithKDCM1}
  T_1 = T \sqrt{
      K \left({V_O -V_I\over V_O}\right)
  }.
\end{equation}
and CCM as:
\begin{equation}
  \label{eqn:T1calcwithKCCM1}
  T_1 = {T\over 4} \left(
   1 - \sqrt{
           1 - { 16 K V_I \over  V_O}
            } \right).
\end{equation}
It can further be shown that the boundary condition of eqn.~\ref{eqnDCMCCMboundary}
can be written as:
\begin{equation}
  \label{eqn:boundarywithK}
  V_O ( 1 - 4K) \ge V_I 
\end{equation}
The feedback loop of Fig.~\ref{fig:feedback} can then be 
used to control the power supply. In Fig.~\ref{fig:feedback}, the power supply output
voltage $V_O$ is measured and compared to a reference voltage $V_{REF}$ to produce
an output voltage error $V_{ERR} = V_O - V_{REF}$.  This error voltage is used by a 
PID controller with dynamics below the input AC frequency $f_{AC}$ to adjust the variable
$K$ to control the output voltage $V_O$.

The variable $K$, is used in the timing generator to generate the inverter timing and
the secondary shorting period $T_1$ twice per sample period $T$.  The timing generator
uses the measured power supply output voltage $V_O$, and a scaled version of the input
rectifier voltage $V_R$ as $V_I = { 1 \over 2 } {N_s \over N_p} V_R$.
Using $K$, $V_I$ and $V_O$, the timing generator evaluates the condition in 
eqn.~\ref{eqn:boundarywithK} and if the result is true, the DCM is selected and
eqn.~\ref{eqn:T1calcwithKDCM1} is used to calculate the time period $T_1$.
Otherwise, the CCM is selected and eqn.~\ref{eqn:T1calcwithKCCM1} is used 
to calculate the time period $T_1$.

\begin{figure}[htb]
  \centering
    \includegraphics[width = 0.475\textwidth]{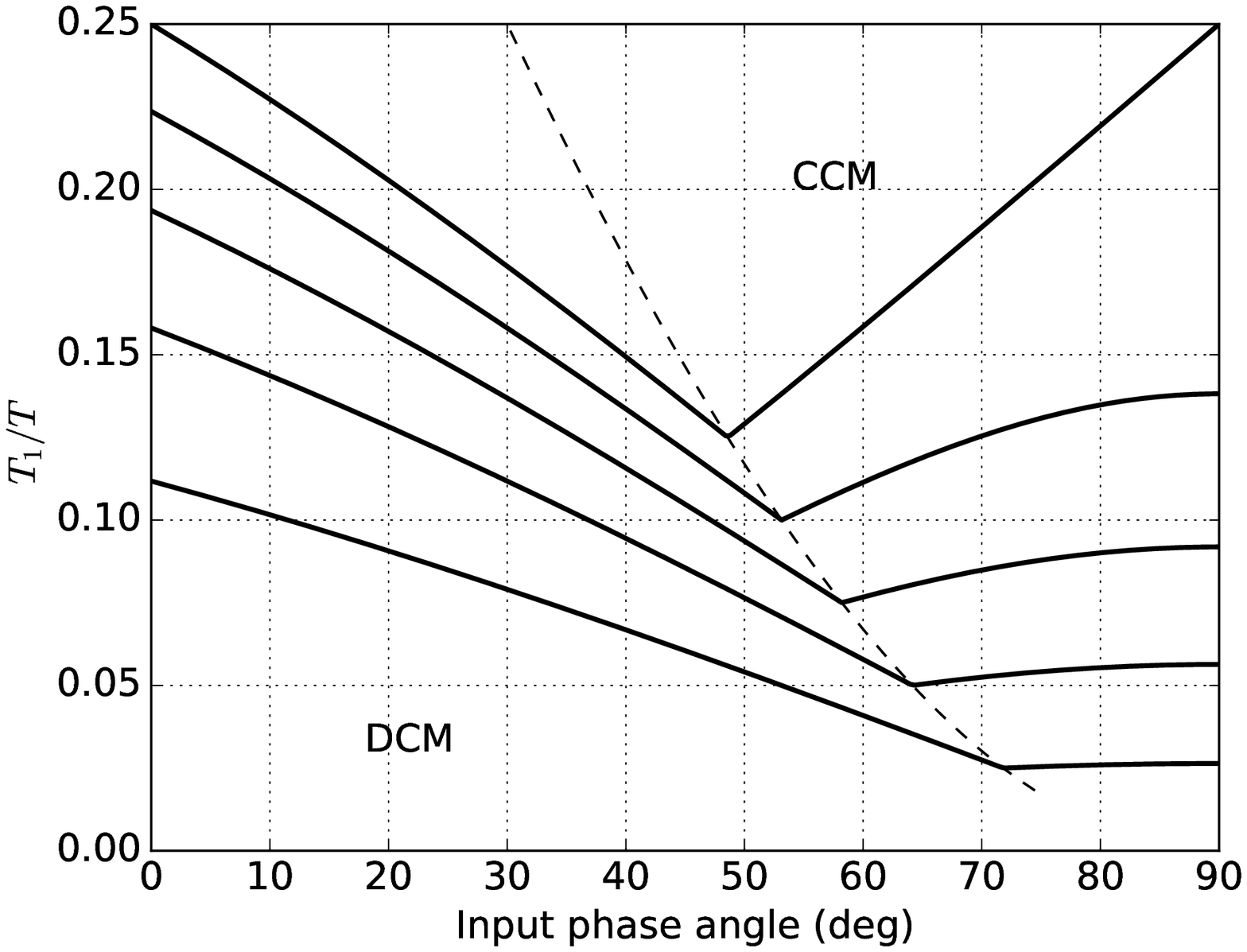}
  \caption{Required timing for $T_1/T$ over one quarter cycle of input sinewave 
       with $G_M$ = 0.2, 0.4, 0.6, 0.8, 1.0 times the maximum $G_M$. }
  \label{fig:T1calceg}
\end{figure}

Fig.~\ref{fig:T1calceg} shows a plot of the $T_1$ time as a fraction of the
switching period $T$ for the first quarter of the input sine wave, under
a range of load conditions from 20\% to full load.
 Under full load ($G_M$ at its maximum value), the circuit operates in discontinuous
conduction mode for just over half the time, with the calculated shorting period $T_1$
reducing from an initial value of $T/4$.  However, after switching to
the continuous operating mode, the calculated shorting period $T_1$ rises
again.  The fraction of time spent in the continuous conduction mode reduces as
the load level drops.

The calculations for shorting period $T_1$ need to be calculated in real-time
by the controller at a rate considerably higher than the input frequency $f_{AC}$,
though not necessarily at the full $f_S$ rate.  Eqn.~\ref{eqn:boundarywithK},
eqn.~\ref{eqn:T1calcwithKDCM1} and eqn.~\ref{eqn:T1calcwithKCCM1}  can be readily
implemented in a modern microcontroller, DSP or FPGA at the required data rates.

\section{Example Prototype Design\label{sec:egdesign}}
Design of a converter using the proposed technique for a given power level $P$, 
an output voltage $V_O$ and a maximum input voltage $V_{AC}$ is based on 
selecting a desired switching frequency $f_s$, calculating the maximum
transformer turns ratio ${N_s \over N_p}$, and using eqn.~\ref{eqn:maxpowercalc}
to determine the maximum allowed value of leakage inductance.

Consider the design of a 300W 50V power supply with a 240Vrms AC 50Hz source or
$P = 300W$, $V_O = 50V$ and $V_{AC}\le 240Vrms$.  A switching frequency
of  $f_s = 50kHz$ is chosen for this example.

Operation in boost mode requires the peak transformer output voltage
to be not greater than the desired output voltage and thus
$$
    {N_s \over N_p}  \le {2 V_O \over \sqrt{2} V_{AC} }
$$
or $ {N_s \over N_p} \le 0.295$.  Choosing $ {N_s \over N_p} = {6 \over 22}$, then rearranging eqn.~\ref{eqn:maxpowercalc} to:
$$
  L_L  \le { V_{AC}  {N_s \over N_p} V_O \over 32 \sqrt{2} f_s P }
$$
imposes the requirement that $ L_L \le 4.82 uH$.
With these parameters, the maximum transformer peak current can be calculated
from eqn.~\ref{eqn:Ipeakcalc} as 26.0 A.


\subsection{Transformer Design}

\begin{figure}[htb]
  \centering
    \includegraphics[width = 0.5\textwidth]{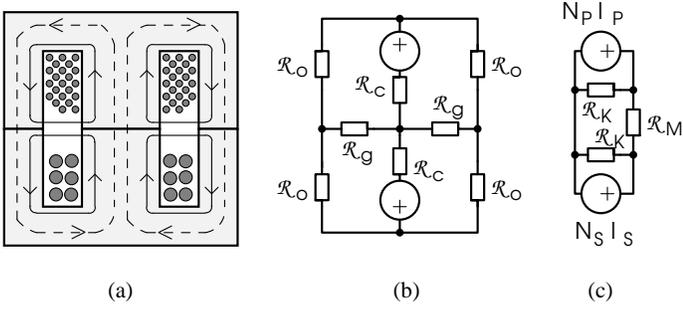}
  \caption{Ecore transformer construction with flux paths for the leakage (solid) and
 magnetizing (dashed) inductances and corresponding magnetic models.}
  \label{fig:txfmerArch}
\end{figure}

The transformer design for the prototype is based on using two
E-Cores with the primary winding on one core and the secondary
winding on the other core as in Fig.~\ref{fig:txfmerArch}(a). 
This results in a transformer 
with good magnetizing inductance and a usable amount of
leakage inductance.   The transformer design requirements are
to provide the specified leakage inductance and maximum  
magnetizing inductance, while preventing core saturation under peak
operating currents. Fig.~\ref{fig:txfmerArch}(b) shows a magnetic 
reluctance model for the transformer with the leakage being modeled
by the air gap reluctance $\mathcal{R}_g$ between the central core and
the outer legs. The central core reluctance is $\mathcal{R}_C$ and the reluctance of
each top and side halves is $\mathcal{R}_O$   The final reluctance model,  Fig.~\ref{fig:txfmerArch}(c)
can then be used with the equations:
$$
  \mathcal{R}_M = 2\mathcal{R}_C + \mathcal{R}_O 
    + 2{(\mathcal{R}_C+ {\mathcal{R}_O\over 2})^2 \over \mathcal{R}_g},
$$
$$
\mathcal{R}_K = \mathcal{R}_C + {\mathcal{R}_O\over 2} + {R}_g.
$$
The resulting magnetizing inductance (from the primary side) is
$L_P ={ N_P^2\over \mathcal{R}_M }$ and total leakage inductance 
(from the secondary side) is $ L_L = 2 N_S^2/  \mathcal{R}_K$.

The peak core magnetic flux $\Phi_{MAX}$ is then the sum of the magnetizing
flux $ N_P I_M \over \mathcal{R}_M$ and the leakage flux $
I_{PS} N_S \over \mathcal{R}_K$, with $I_M$ being the peak primary magnetizing
current and  $I_{PS}$ being the peak secondary current.
The peak core magnetic flux density with a core effective area of $A_e$ is 
then:
\begin{equation}
  \label{eqn:Bsateq}
 B_{MAX} = {N_P I_M \over \mathcal{R}_M A_e}
    + { I_{PS} N_S \over \mathcal{R}_K A_e}
\end{equation}

The prototype transformer consists of a pair of E42/21/15 cores, 3C90 ferrite
material, with 22 turns on the primary and 6 turns on the secondary.  The 
corresponding magnetizing inductance (from the primary side) and
total leakage inductance (from the secondary side) is calculated as:
$$
 L_M = 1800\mu H \quad\quad\mbox{ and }\quad\quad
 L_L = 4.7\mu H
$$
and the actual measured values are listed in Table~\ref{tab:measuredxfmr}.
With a $240V_{rms}$ AC input, the magnetizing current is $\approx \pm 0.47A$
and with a peak secondary current of
$26A$, the peak core magnetic flux density based on 
equation~\ref{eqn:Bsateq} is
$$
  B_{MAX} = 0.211 T + 0.0562 T ~\approx 0.27 T.
$$

This value is less than the saturation magnetic flux density of $0.32T$, and
only occurs at the peak of the input AC waveform.
Notice that the contribution from the leakage flux is small due to the high
reluctance of the leakage flux path as it passes through the air gap.  
This example shows that the addition of the leakage inductance does 
not severely impact on the size of transformer in terms of core saturation.


\subsection{Prototype Construction and Measurements\label{sec:prototypemeasure}}
\begin{figure*}[htb]
  \centering
    \includegraphics[width = 0.75\textwidth]{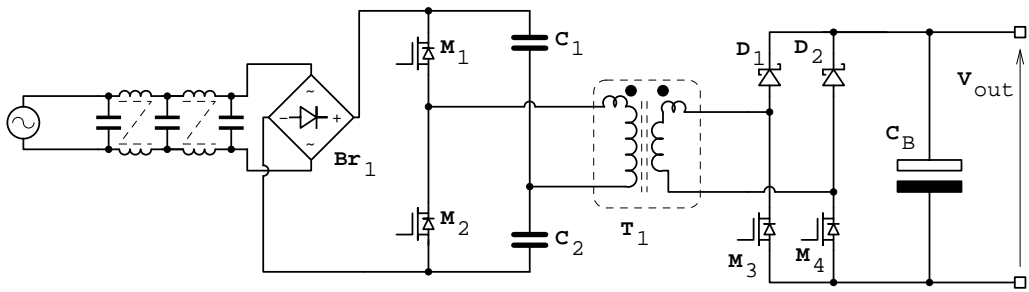}
  \caption{Prototype power electronics schematic.}
  \label{fig:practice}
\end{figure*}

The power electronics schematic of the prototype power supply is shown in 
Fig.~\ref{fig:practice}.  A bridge rectifier $\mbox{Br}_1$ is used to
rectify the input AC voltage and a half bridge inverter based on MOSFETs M1 
and M2 drive the transformer.
The prototype is based on a semi synchronous rectifier for the
output rectifier \cite{ref:SoftSwitchDCDC}.  In this case, two MOSFET transistors M3 and M4 are
used to perform the secondary shorting during the $T_1$ period by turning
on both transistors. During the remaining time, only one of
the transistors is left switched on to perform the rectification 
function in conjunction with one of the Schottky diodes, D1 or D2.

The measured parameters of the constructed prototype transformer
are listed in Table~\ref{tab:measuredxfmr}, together with the 
other power components used in the prototype power supply.   
  \begin{table}[htb]
      \begin{center}
         \caption{Prototype component values and specification.}
         \begin{tabular}{|c|c|}
            \hline
Component/Parameter  &  Specified/Measured \\
            \hline
            \hline
 EMC filter & $3\times$ 100nF Capacitor \\
            & $2\times$ 5A 7mH Common Mode Choke \\
&\\
 Rectifier $\mbox{Br}_1$ & BU15065S-E3/45, 15A 600V \\
                       & Bridge Rectifier \\
MOSFETS M1, M2 &  C3M0280090D SiC, 11.5 A, 900 V\\
&\\
Capacitors C1, C2 &  1uF 250V Film Capacitor (PET) \\
&\\
  Transformer  & $2\times$ Ecore 42mm/21mm/15mm, 3C90  \\
 Turns Ratio &   22:6 \\
 Primary Inductance &   2.0 mH \\
 Secondary Inductance &  150 uH \\
Secondary Leakage & 4.0 uH \\
  Inductance &   \\
&\\
Schottky Diodes D1, D2 & MBR4060  60V 40A \\
MOSFETS M3, M4 & IRFB4110 100V 120A \\
&\\
Bulk Capacitor &  6000uF (6$\times$1000uF) 63V \\
&\\
 Control Processor & LPC1114 ARM M0  \\
            \hline
         \end{tabular}
         \label{tab:measuredxfmr}
      \end{center}
   \end{table}

Control of the prototype is implemented in firmware on 
a modern 32bit microcontroller. An internal 10 bit ADC 
measures the input rectified voltage and bulk capacitor 
output voltage.  The microcontroller performs the calculations 
for the PID output voltage control and the timing calculations as per 
section~\ref{sec:psucontrol} with the timing signals generated 
with on-chip PWM generators.

\begin{figure}[htb]
  \centering
    \includegraphics[width = 0.5\textwidth]{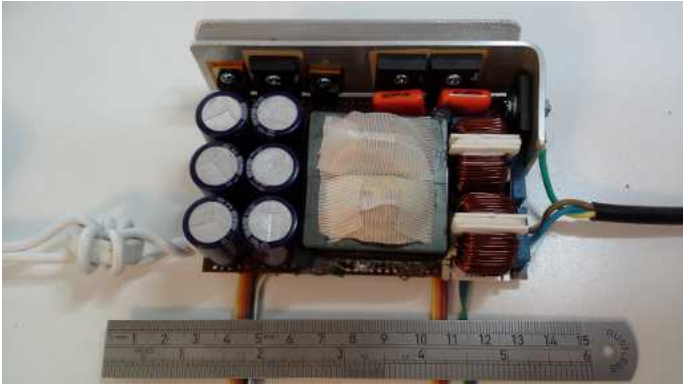}
  \caption{Prototype power supply.}
  \label{fig:photo}
\end{figure}

\begin{figure}[htb]
  \centering
    \includegraphics[width = 0.5\textwidth]{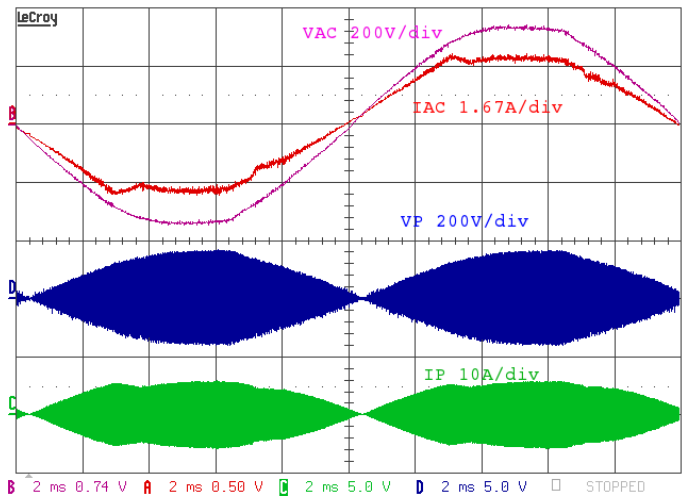}
  \caption{Measured waveforms at 300W operation over a complete line cycle, 
   input AC voltage (VAC) and current (IAC) and 
    transformer primary voltage (VP) and current (IP). }
  \label{fig:waveform1}
\end{figure}

\begin{figure}[htb]
  \centering
    \includegraphics[width = 0.5\textwidth]{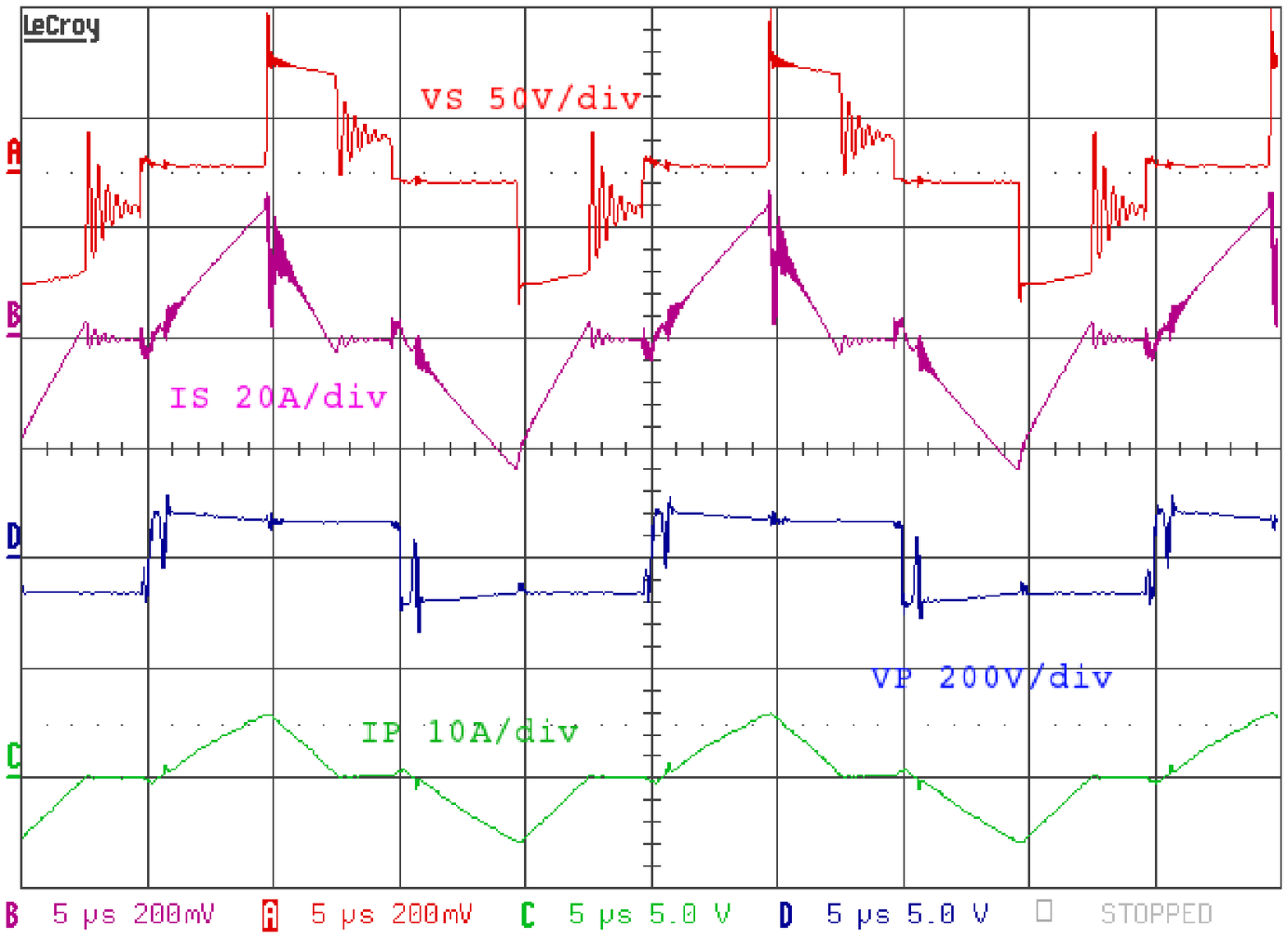}
  \caption{Measured waveforms at 300W operation, 1.5ms from zero crossing and operating in DCM mode.
    Transformer primary voltage (VP) and current (IP) and secondary voltage (VS) and current (IS). }
  \label{fig:waveform2}
\end{figure}

\begin{figure}[htb]
  \centering
    \includegraphics[width = 0.5\textwidth]{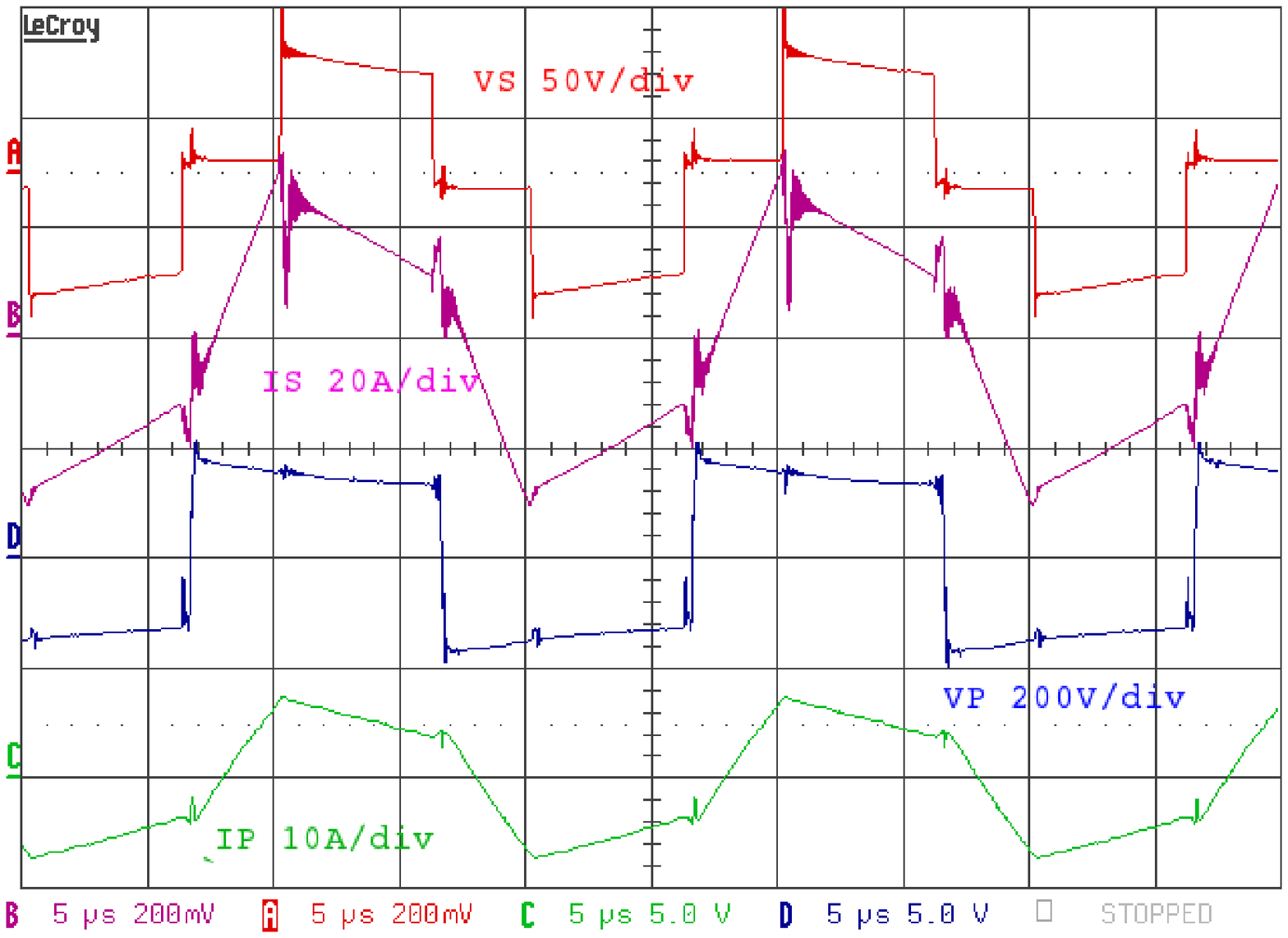}   
  \caption{Measured waveforms at 300W operation, 5ms from zero crossing and operating in CCM mode.
    Transformer primary voltage (VP) and current (IP) and secondary voltage (VS) and current (IS). }
  \label{fig:waveform3}
\end{figure}

Fig.~\ref{fig:photo} shows an image of the prototype power supply
and its measured performance at 300W output power is given 
in Table~\ref{tab:measuredpf}.
Measured waveforms of the line input voltage and current and transformer
primary voltage and current are shown in Fig.~\ref{fig:waveform1}
over a full line cycle.  Zoomed in waveforms of the transformer
primary voltage and current and secondary voltage and current
are shown in Fig.~\ref{fig:waveform2} (DCM) and Fig.~\ref{fig:waveform3} (CCM)
and confirm the desired operation.
The effect of finite values of bus capacitors C1 and C2 can be seen in 
the primary voltage waveform of Fig.~\ref{fig:waveform3} as a droop
in the voltage rather than an ideal square wave.

\begin{figure}[htb]
  \centering
    \includegraphics[width = 0.45\textwidth]{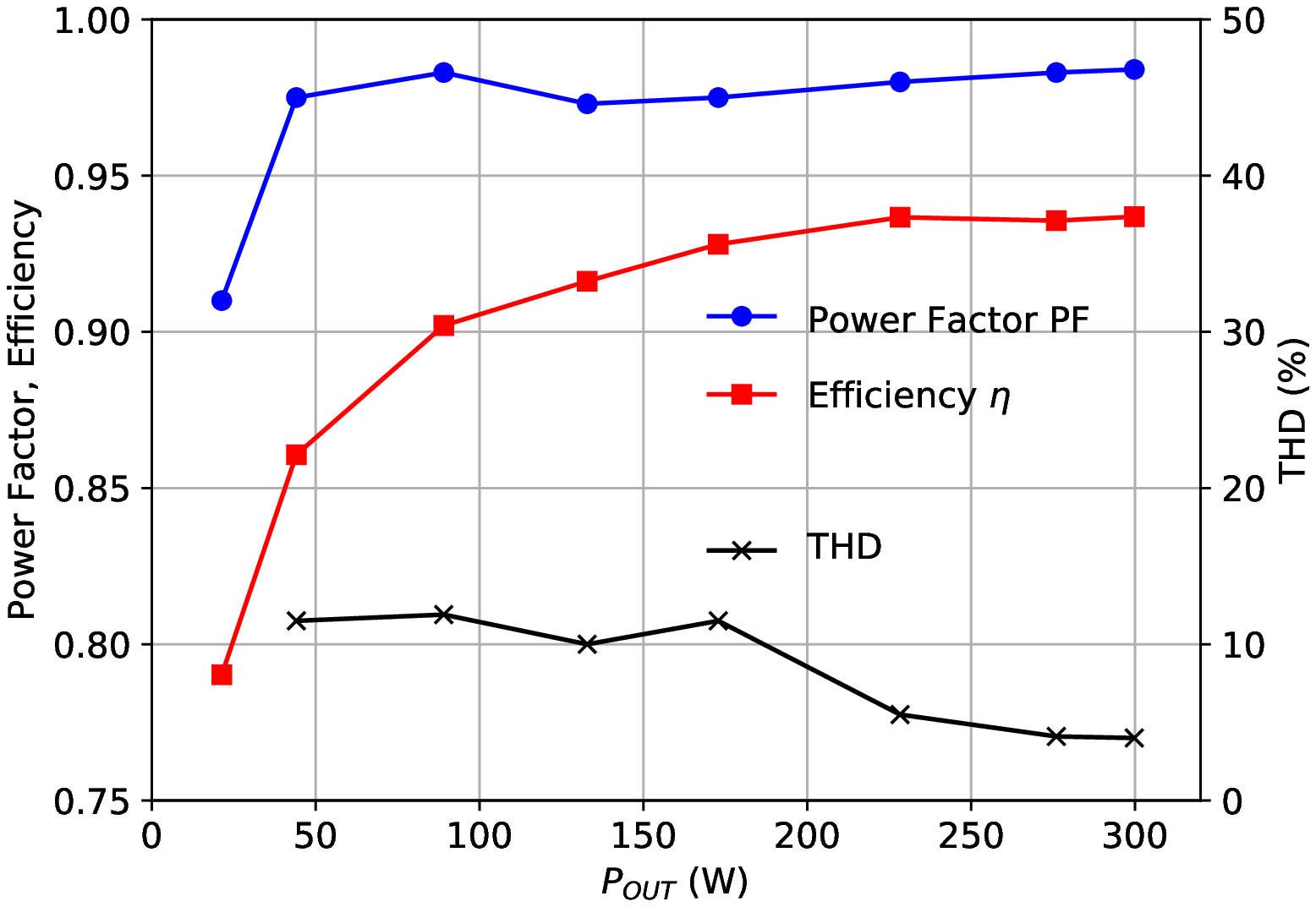}
  \caption{Measured efficiency and power factor for the prototype.}
  \label{fig:measeffPF}   
\end{figure}

The measured efficiency, power factor and total harmonic distortion are 
plotted against output power in Fig.~\ref{fig:measeffPF}.

  \begin{table}[htb]
      \begin{center}
         \caption{Measured performance for the experimental prototype.}
         \begin{tabular}{|c|c|}
            \hline
Parameter &  Measured Value \\
            \hline
            \hline
Switching Frequency & 50 kHz \\
Input Voltage & 237.1Vrms \\
Input Current & 1.375Arms \\
Input Power   & 320W \\
Power Factor  & 0.98 \\
&\\
Output Voltage & 49.8V \\
Output Voltage Ripple & $ 3.8V_{pp}$ (i.e $\pm 3.8\%$) \\    
Output Current & 6.02A \\
Output Power   & 300W \\
&\\
Efficiency &  93.7 \% \\
&\\
THD &  $4.1\%$,   \\

            \hline
         \end{tabular}
         \label{tab:measuredpf}
      \end{center}
   \end{table}

\begin{figure}[htb]
  \centering
    \includegraphics[width = 0.45\textwidth]{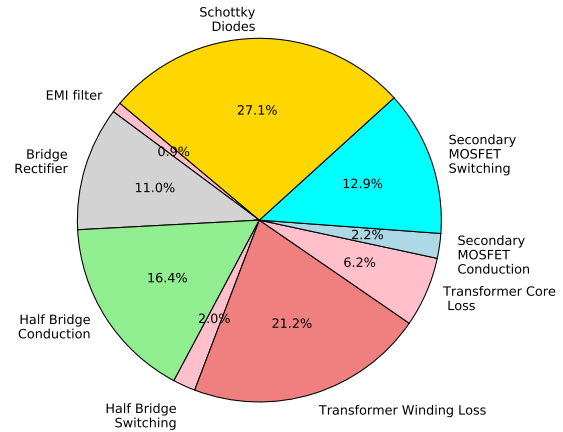}
  \caption{Calculated loss contributions from the circuit elements.}
  \label{fig:effpie.eps}
\end{figure}

\begin{figure}[htb]
  \centering
    \includegraphics[width = 0.4\textwidth]{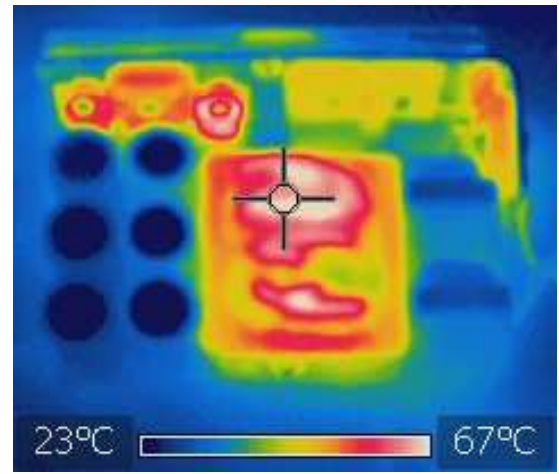}  
  \caption{Thermal Image of prototype at 300W output power (Spot temperature 62.1 deg C).}
  \label{fig:thermalphoto}
\end{figure}

The efficiency of the converter is determined by the component losses and an
analytical model of the relative contributions results in the distribution 
shown in Fig.~\ref{fig:effpie.eps}.  The largest contributions are given by
the Schottky diodes conduction losses and the transformer core and winding 
losses.  
Fig.~\ref{fig:thermalphoto} shows an infra red thermal image of the prototype
showing the highest spot temperature on the windings of the transformer.
Switching losses dominate in the secondary MOSFET switches and the use of 
faster switches such as GAN devices or soft switching techniques might be 
useful here particularly if a higher frequency of operation is adopted.

  
The prototype design is based on a fixed line input voltage
and a fixed switching frequency of 50kHz.  Further work could
involve extending the architecture to operate as an universal
supply over wide input voltage operation.  While the existing
design will work at lower input voltages, the output power
is limited by the fixed switching frequency and leakage inductance
value as per equation~\ref{eqn:maxpowercalc}.
For a given power level, universal input voltage support could be
addressed by adopting variable frequency operation with lower input
voltages supported by reducing the switching frequency.

\subsection{Comparison with existing designs}
A key motivation for the work reported in this paper is the reduction of size
and weight of the converter with a key component of these being the
magnetic elements.
Table~\ref{tab:pfccmp} provides a comparison of the prototype developed with a range
of PFC isolated supplies described in the literature.  Flyback type architectures
dominate at lower power levels (below 100W) with two stage architectures accounting
for higher power levels.  The prototype based on the proposed architecture in this
paper compares well with the other designs and industry expectations \cite{ref:effvalues}, and 
should provide a useful topology for PFC converter designers.

  \begin{table*}[htb]
      \begin{center}
{\scriptsize 
         \caption{Comparison with other isolated PFC topologies.}
         \begin{tabular}{|l|c|c|c|c|c|c|c|c|}
            \hline
		 Ref:                 &  Topology & Switching &  Peak &  Universal  & Weight of      & No. Switches (S) & Peak   & Power per\\
		     &           & Frequency & Power  & Voltage    & Magnetic Cores & No. Diodes (D) &Efficiency & Core Weight \\
            \hline
            \hline
 \cite{ref:pfcflyback}    & Flyback & 65kHz     &    60W   &   Yes      &        31g   & 1 S, 5 D   &   89.5\%  &  1.94 W/g\\

 \cite{ref:pfccmp1} & Flyback &  ( $\ge$ 44kHz)  &    60W  &    Yes     &         36g   & 1 S, 5 D   &  91\%   & 1.67 W/g  \\

 \cite{ref:pfccmp3} & Flyback & 40kHz      &    72W   &    No      &        36g    & 2 S, 2 D  &   90\% & 2.00 W/g \\

 \cite{ref:pfccmp2} & Boost-Flyback & 50kHz     &    96W  & Yes    &        120g  &  1 S, 7 D & 90.6\%  & 0.80 W/g \\   

 \cite{ref:pfccmp5} & Quasi-Active Flyback & 75kHz      &   100W   &   Yes      &        57g   & 1 S, 10 D   &   93\%  & 1.75 W/g\\  

 \cite{ref:PostxfmrCbulkandBoostInductorSize} & Resonant LLC / Boost & 90-450kHz & 250W & Yes & 92g &  8 S , 4 D &94.5\% & 2.72 W/g \\   

 \cite{ref:pfccmp6} & Boost / Resonant LLC & 100kHz       &  480W &  No & 122g  & 5 S, 4 D & 93.58\% & 3.93 W/g\\  

	 & {\it This work}  & 50kHz &   300W  &     No     &   88g   & 4 S, 6 D   &  93.7\% & 3.41 W/g \\

            \hline
         \end{tabular}
         \label{tab:pfccmp}
   }
      \end{center}
   \end{table*}



\section{Conclusions}

This paper describes an isolated AC/DC power supply using the 
leakage inductance of the isolation transformer to achieve
active power factor correction.  The proposed architecture 
allows for a compact lightweight power supply for power levels
above that of flyback type PFC supplies.
  The principle of operation with two conduction modes is  
described and a timing based control method is developed
for the power factor control.
  A prototype power supply is designed and constructed to
verify experimentally the operating principle. 
Measurements confirm the active power factor correction
functionality with high power factor and low THD.



\begin{thebibliography}{1}
\bibitem{ref:pfcgeneral}
O. Garcia, J. A. Cobos, R. Prieto, P. Alou and J. Uceda, "Single phase power factor correction: a survey," in IEEE Transactions on Power Electronics, vol. 18, no. 3, pp. 749-755, May 2003.

\bibitem{ref:rev2}
B. Singh, B. N. Singh, A. Chandra, K. Al-Haddad, A. Pandey and D. P. Kothari, "A review of single-phase improved power quality AC-DC converters," in IEEE Transactions on Industrial Electronics, vol. 50, no. 5, pp. 962-981, Oct. 2003.

\bibitem{ref:rev1}
M. M. Jovanovic and Y. Jang, "State-of-the-art, single-phase, active power-factor-correction techniques for high-power applications - an overview," in IEEE Transactions on Industrial Electronics, vol. 52, no. 3, pp. 701-708, June 2005.

\bibitem{ref:MohanBook}
N. Mohan, T. M. Undeland and W. P. Robbins, "Power Electronics: Converters, Applications, and Design", 
Wiley, Oct 2002, ISBN: 978-0-471-22693-2 

\bibitem{ref:pfcflyback}
T. Yan, J. Xu, F. Zhang, J. Sha and Z. Dong, "Variable-On-Time-Controlled Critical-Conduction-Mode Flyback PFC Converter," in IEEE Transactions on Industrial Electronics, vol. 61, no. 11, pp. 6091-6099, Nov. 2014.

\bibitem{ref:pfcflybackLED}
D. G. Lamar, M. Arias, A. Rodriguez, A. Fernandez, M. M. Hernando and J. Sebastian, "Design-Oriented Analysis and Performance Evaluation of a Low-Cost High-Brightness LED Driver Based on Flyback Power Factor Corrector," in IEEE Transactions on Industrial Electronics, vol. 60, no. 7, pp. 2614-2626, July 2013.

\bibitem{ref:pfccmp1}
C. Zhao, J. Zhang and X. Wu, "An Improved Variable On-Time Control Strategy for a CRM Flyback PFC Converter," in IEEE Transactions on Power Electronics, vol. 32, no. 2, pp. 915-919, Feb. 2017.

\bibitem{ref:pfccmp3}
J. W. Shin, S. J. Choi and B. H. Cho, "High-Efficiency Bridgeless Flyback Rectifier With Bidirectional Switch and Dual Output Windings," in IEEE Transactions on Power Electronics, vol. 29, no. 9, pp. 4752-4762, Sept. 2014.

\bibitem{ref:pfcfly100W36V}
C. p. Tung and H. S. h. Chung, "A flyback AC/DC converter using power semiconductor filter for input power factor correction," 2016 IEEE Applied Power Electronics Conference and Exposition (APEC), Long Beach, CA, 2016, pp. 1807-1814.

\bibitem{ref:PostxfmrCbulkandBoostInductorSize}
L. Gu, W. Liang, M. Praglin, S. Chakraborty and J. M. Rivas Davila, "A Wide-Input-Range High-Efficiency Step-down Power Factor Correction Converter Using Variable Frequency Multiplier Technique," to appear in IEEE Transactions on Power Electronics 2018

\bibitem{ref:pfcinrush}
 K. Mino, H. Matsumoto, S. Fujita, Y. Nemoto, D. Kawasaki, R. Yamada and N. Tawada,
"Novel bridgeless PFC converters with low inrush current stress and high efficiency," 
The 2010 International Power Electronics Conference - ECCE ASIA -, Sapporo, 2010, pp. 1733-1739.

\bibitem{ref:pfcinrush2}
M. Alam, W. Eberle and N. Dohmeier, "An inrush limited, surge tolerant hybrid resonant bridgeless PWM AC-DC PFC converter," 2014 IEEE Energy Conversion Congress and Exposition (ECCE), Pittsburgh, PA, 2014, pp. 5647-5651.
\bibitem{ref:pfcflybackVOT}
A. H. Memon, K. Yao, Q. Chen, J. Guo and W. Hu, "Variable-On-Time Control to Achieve High Input Power Factor for a CRM-Integrated Buck-Flyback PFC Converter," in IEEE Transactions on Power Electronics, vol. 32, no. 7, pp. 5312-5322, July 2017.

\bibitem{ref:pfccmp5}
H. S. Athab, D. Dah-Chuan Lu, A. Yazdani and B. Wu, "An Efficient Single-Switch Quasi-Active PFC Converter With Continuous Input Current and Low DC-Bus Voltage Stress," in IEEE Transactions on Industrial Electronics, vol. 61, no. 4, pp. 1735-1749, April 2014.

\bibitem{ref:pfccmp2}
S. W. Lee and H. L. Do, "A Single-Switch AC-DC LED Driver Based on a Boost-Flyback PFC Converter With Lossless Snubber," in IEEE Transactions on Power Electronics, vol. 32, no. 2, pp. 1375-1384, Feb. 2017.

\bibitem{ref:pfccmp6}
J. I. Baek, J. K. Kim, J. B. Lee, H. S. Youn and G. W. Moon, "A Boost PFC Stage Utilized as Half-Bridge Converter for High-Efficiency DC-DC Stage in Power Supply Unit," in IEEE Transactions on Power Electronics, vol. 32, no. 10, pp. 7449-7457, Oct. 2017.

\bibitem{ref:cntlleakageinduct}
M. A. Bakar and K. Bertilsson, "An improved modelling and construction of power transformer for controlled leakage inductance," 2016 IEEE 16th International Conference on Environment and Electrical Engineering (EEEIC), Florence, 2016, pp. 1-5.

\bibitem{ref:GANgeneral}
R. Mitova, R. Ghosh, U. Mhaskar, D. Klikic, M. X. Wang and A. Dentella, "Investigations of 600-V GaN HEMT and GaN Diode for Power Converter Applications," in IEEE Transactions on Power Electronics, vol. 29, no. 5, pp. 2441-2452, May 2014.

\bibitem{ref:SiCHardSwitched}
A. M. Abou-Alfotouh, A. V. Radun, H. R. Chang and C. Winterhalter, "A 1-MHz hard-switched silicon carbide DC-DC converter," in IEEE Transactions on Power Electronics, vol. 21, no. 4, pp. 880-889, July 2006.

\bibitem{ref:SiCHardSwitched2}
J. S. Glaser, J. J. Nasadoski, P. A. Losee, A. S. Kashyap, K. S. Matocha, J. L. Garrett and L. D. Stevanovic,
"Direct comparison of silicon and silicon carbide power transistors in high-frequency hard-switched applications," 2011 Twenty-Sixth Annual IEEE Applied Power Electronics Conference and Exposition (APEC), Fort Worth, TX, 2011, pp. 1049-1056.

\bibitem{ref:SoftSwitchDCDC}
H. Wu, Y. Lu, T. Mu and Y. Xing, "A Family of Soft-Switching DC-DC Converters Based on a Phase-Shift-Controlled Active Boost Rectifier," in IEEE Transactions on Power Electronics, vol. 30, no. 2, pp. 657-667, Feb. 2015.



%
%
%
%
%
\bibitem{ref:effvalues}
C. A. Quinn and D. B. Dalal, "Empowering the electronics industry: a power technology roadmap," in CPSS Transactions on Power Electronics and Applications, vol. 2, no. 4, pp. 306-319, December 2017.

\end{thebibliography}
\end{document}